\documentclass[10pt,twocolumn,twoside,letterpaper]{IEEEtran}

\usepackage{graphicx}
\usepackage{cite}
\usepackage[cmex10]{amsmath}
\usepackage{amssymb}

\usepackage{tikz}
\usepackage{pgfplots}
\pgfplotsset{compat=1.3}% <-- moves axis labels near ticklabels (respects tick label widths)
\usetikzlibrary{arrows,snakes,shapes}

\DeclareMathAlphabet{\mathbit}{OML}{cmr}{bx}{it}

\DeclareMathOperator{\E}{E}
\DeclareMathOperator{\T}{T}

\DeclareMathOperator{\fieldR}{\mathbb{R}}

\newcommand{\ve}[1]{\boldsymbol{#1}}

\newcommand{\exdi}[2]{\E_{#1} \left[#2\right]}

\renewcommand{\exp}[1]{\operatorname{exp}\left(#1\right)}

\newtheorem{theorem}{Theorem}

\title{Measurement-driven Quality Assessment of Nonlinear Systems by Exponential Replacement}

\author{
\IEEEauthorblockN{Manuel~Stein, Josef A. Nossek and Kurt Barb\'{e}\IEEEauthorrefmark{1}}
\\
\IEEEauthorblockA{Institute for Circuit Theory and Signal Processing (NWS), Technische Universit\"at M\"unchen, Germany} %D--80333
\\   \IEEEauthorblockA{\IEEEauthorrefmark{1}Research Team Stochastics (STOX), Dept. Mathematics (DWIS), Vrije Universiteit Brussel, Belgium}    %D--82234
\\ Email: manuel.stein@tum.de, josef.a.nossek@tum.de, kurt.barbe@vub.ac.be}

\begin{document}
\maketitle
\begin{abstract}
We discuss the problem how to determine the quality of a nonlinear system with respect to a measurement task. Due to amplification, filtering, quantization and internal noise sources physical measurement equipment in general exhibits a nonlinear and random input-to-output behaviour. This usually makes it impossible to accurately describe the underlying statistical system model. When the individual operations are all known and deterministic, one can resort to approximations of the input-to-output function. The problem becomes challenging when the processing chain is not exactly known or contains nonlinear  random effects. Then one has to approximate the output distribution in an empirical way. Here we show that by measuring the first two sample moments of an arbitrary set of output transformations in a calibrated setup, the output distribution of the actual system can be approximated by an equivalent exponential family distribution. This method has the property that the resulting approximation of the statistical system model is guaranteed to be pessimistic in an estimation theoretic sense. We show this by proving that an equivalent exponential family distribution in general exhibits a lower Fisher information measure than the original system model. With various examples and a model matching step we demonstrate how this estimation theoretic aspect can be exploited in practice in order to obtain a conservative measurement-driven quality assessment method for nonlinear measurement systems.
\end{abstract}
\begin{keywords}
nonlinear systems, Fisher information, Cram\'er-Rao lower bound, exponential family, Saleh model, Rician model, regression, Wiener system, measurement uncertainty
\end{keywords}
\section{Motivation}
The characterization of nonlinear systems and the development of appropriate processing algorithms forms a problem for engineering tasks like signal processing or system identification while the increasing demand on low-cost, energy-efficient and fast measurement devices makes it inevitable to operate such systems outside their linear regime. In order to provide high processing accuracy under these circumstances, it is important to investigate nonlinear models and to develop generic approaches and methods for such kind of problems. The key challenge for efficient solutions to nonlinear measurement problems lies in the fact that access to an appropriate model $p(z;\theta)$ is required. In theory, this probabilistic description establishes the statistical relationship of the system parameter $\theta\in\fieldR$ and the output measurement $z\in\fieldR$. This allows to formulate efficient signal processing algorithms and to describe the achievable performance by analytical tools. However, in practice the system model $p(z;\theta)$ is rarely known exactly and therefore has to be established by numerical calculations or approximated by physical measurements.

If the analytical form of the system model $p(z;\theta)$ is known but to complicated to work with, one can resort to an direct approximation of the distribution function like performed in \cite{Medawar13} for the Rician model. When the nonlinear output is a parametric but deterministic transformation $f(x;\theta)$ of an intermediate variable $x\in\fieldR$ with known statistical model $p(x;\theta)$, the input-to-output relation $z=f(x;\theta)$ can be represented by a Taylor expansion \cite{Oehlert92}. If the input to the nonlinear system is Gaussian, the model can be approximated through a Bussgang decomposition \cite{Rowe82}, where the idea is to use the best linear approximation of the nonlinearity and to characterize the second moment of the resulting error. In the situation where the output model is completely unknown one has to use empirical methods like histograms \cite{Barbe14}. In the case that the system parameter only modulates the output mean and leaves the variance constant, it is possible to use a pessimistic replacement strategy by assuming an equivalent Gaussian model. This becomes possible as for such a situation the Gaussian model is the least favorable distribution \cite{Stoica11,Stein14}.

The method of exponential replacement and subsequent model matching that is presented on the following pages can be interpreted as a generalization of the Gaussian replacement strategy. This replacement can be extended to any distribution of the exponential type, where the gathered measurements are used to estimate the sufficient statistics and model matching is used to determine the natural parameters of the selected exponential distribution. By using the Cauchy-Schwarz inequality we show that such an replacement $\tilde{p}(z;\theta)$ in general exhibits a lower Fisher information measure than the original system model $p(z;\theta)$. We discuss how to match the exponential replacement model $\tilde{p}(z;\theta)$ in an estimation theoretic sense by maximizing its Fisher information. The result of the presented procedure is an approximation of the statistical system model which can be obtained in a measurement-driven way and is guaranteed to be conservative with respect to the possible accuracy that can be obtained by gathering data from $p(z;\theta)$ to measure the parameter $\theta$ without bias. In order to demonstrate the useful character of the exponential replacement, we show how to evaluate the estimation theoretic quality of three nonlinear systems by calibrated measurements at the system output. We use the Saleh model \cite{Saleh81} to simulate an amplification device with saturation effects and unknown input power level. As a second example we use the Rician model with unknown distance parameter, which is of interest in wireless communications \cite{Tepe03} and biomedical applications like fMRI \cite{Barbe11}. Finally, we discuss a polynomial input-to-output relation of cubic order with unknown regression parameter which exhibits the properties of a Wiener type of nonlinear system \cite{Bai03}.
\section{Problem Formulation}
Consider the problem of measuring the system parameter $\theta$ by taking measurements at the output $z$ of a system. The system is represented by a parametrized probability density or mass function $p(z;\theta)$, with random variable $z\in\mathcal{Z}$ and a deterministic system parameter $\theta\in\Theta$. $\mathcal{Z}\subset\fieldR$ is the support of the random output $z$ and $\Theta\subset\fieldR$ the parameter space of $\theta$. Throughout our discussion we assume that all integrands are absolutely integrable on the support $\mathcal{Z}$. All density or mass functions $p(z;\theta)$ exhibit regularity and are differentiable with respect to the parameter $\theta$. For such systems the Fisher information measure is defined by \cite{Kay93} 
\begin{align}
F(\theta)&=\int_{\mathcal{Z}} \bigg(\frac{\partial \ln{p(z;\theta)}}{\partial\theta}\bigg)^2 p(z;\theta)  {\rm d}z.
\label{measure:fisher}
\end{align}
For any unbiased estimator
\begin{align}
\exdi{\ve{z};\theta}{\hat{\theta}(\ve{z})-\theta}=0
\end{align}
using $N$ independent samples, here denoted by $\ve{z}\in\fieldR^N$, the accuracy is limited by the classical Cram\'er-Rao lower bound (CRLB) \cite{Rao45,Cram46}
\begin{align}\label{bound:crlb}
\exdi{\ve{z};\theta}{\big(\hat{\theta}(\ve{z})-\theta\big)^2}\geq \frac{1}{NF(\theta)}.
\end{align}
Note that if $N$ is sufficiently large and $\hat{\theta}(\ve{z})$ is an efficient estimator, the statement \eqref{bound:crlb} holds with equality. Therefore, the Fisher information measure \eqref{measure:fisher} provides a mathematical tool in order to evaluate the asymptotic estimation theoretic quality of the model $p(z;\theta)$. The higher the actual value of the information measure $F(\theta)$, the higher the accuracy that can be expected while estimating $\hat{\theta}(\ve{z})$. However, calculation of \eqref{measure:fisher} requires an analytic description of the log-likelihood function $\ln p(z;\theta)$ and its derivative $\frac{\partial \ln p(z;\theta)}{\partial \theta}$, referred to as the score function. Further, the computation of the integral in \eqref{measure:fisher} can be demanding. This motivates the problem of finding approximations for $F(\theta)$ which has found our attention for the design of quantized signal processing systems \cite{SteinWCL14} and allows to deduce a pessimistic but highly tractable replacement $\tilde{p}(z;\theta)$ for the original model $p(z;\theta)$ \cite{Stein14}.
\section{Exponential Replacement}
Our approach follows the idea of replacing the original system $p(z;\theta)$ by an equivalent distribution $\tilde{p}(z;\theta)$ which belongs to the exponential family.
\subsection{Exponential Family}
An exponential family with a single parameter $\theta$ is a set of probability density or mass functions, which can be factorized such that the form
\begin{align}\label{def:exp:family}
p(z;\theta)=\exp{\sum_{l=1}^{L} w_l(\theta) t_l(z) - \lambda(\theta)+\kappa(z)}
\end{align}
is obtained. Within this factorization $w_l(\theta)$ is the $l$-th natural parameter, $t_l(z)$ is the associated sufficient statistic, $\lambda(\theta)$ is the log-normalizer and $\kappa(z)$ is a carrier measure. The log-likelihood function of an exponential family is given by
\begin{align}
\ln p(z;\theta) &= \sum_{l=1}^{L} w_l(\theta) t_l(z) - \lambda(\theta)+\kappa(z),
\end{align}
while the score function attains the structure
\begin{align}\label{exp:score}
\frac{\partial \ln p(z;\theta)}{\partial \theta} &= \sum_{l=1}^{L} \frac{\partial w_l(\theta)}{\partial \theta} t_l(z) - \frac{\partial\lambda(\theta)}{\partial \theta}.
\end{align}
\subsection{Exponential Replacement}
We substitute the original system $p(z;\theta)$ by $\tilde{p}(z;\theta)$ and assume that the score function of the replacement model $\tilde{p}(z;\theta)$ exhibits an exponential factorization
\begin{align}\label{approx:score}
\frac{\partial \ln \tilde{p}(z;\theta)}{\partial \theta} &= \sum_{l=1}^{L} \beta_l(\theta) \phi_l(z) - \alpha(\theta),
\end{align}
where $\beta_l(\theta)$ and $\alpha(\theta)$ are weighting factors which depend on $\theta$. The functions $\phi_l(z)$ are arbitrary transformations of the output data $z$.
\subsection{Fisher Information Bound}
It can be shown that the approximation \eqref{approx:score} provides a pessimistic system model in an estimation theoretic sense. This result is obtained by invoking the Cauchy-Schwarz inequality with the exponential score function \eqref{approx:score}.
\begin{theorem}[Fisher Information Bound]
For a probability density or mass function $p(z;\theta)$, a set of $L$ output transformations $\phi_l(z)$ and arbitrary weighting functions $\beta_l(\theta)$, the Fisher information measure is lower bounded by 
\begin{align}\label{eq:strong:fbound:unopt}
&F(\theta)\geq\notag\\
&\frac{\Big( \sum\limits_{l=1}^{L} \beta_l(\theta)  \frac{\partial \exdi{z;\theta}{\phi_l(z)} }{\partial\theta} \Big)^2}{\exdi{z;\theta}{ \Big( \sum\limits_{l=1}^{L} \beta_l(\theta) \phi_l(z) \Big)^2 } - \Big( \sum\limits_{l=1}^{L} \beta_l(\theta)\exdi{z;\theta}{\phi_l(z)}\Big)^2}.
\end{align}
\end{theorem}
\begin{IEEEproof}
see Appendix \ref{append:fish:bound}.
\end{IEEEproof}
\subsection{Model Matching}
It is possible to optimize the weighting factors $\beta_l(\theta)$ such that the right hand side of the information bound \eqref{eq:strong:fbound:unopt} is maximized. In order to obtain a compact problem formulation, we define the transformation vector
\begin{align}
\ve{\phi} (z)  =  \begin{bmatrix} \phi_1(z) &\phi_2(z) &\ldots &\phi_L(z) \end{bmatrix}^{\T}.
\end{align}
The corresponding weighting functions are summarized by
\begin{align}
\ve{\beta} (\theta) =  \begin{bmatrix} \beta_1(\theta) &\beta_2(\theta) &\ldots &\beta_L(\theta) \end{bmatrix}^{\T}.
\end{align}
Further, we define the mean of the transformation vector
\begin{align}\label{mean:aux:stat}
\ve{\mu}_{\ve{\phi}}(\theta)=\exdi{z;\theta}{\ve{\phi}(z)}
\end{align}
and its covariance
\begin{align}\label{covariance:aux:stat}
\ve{R}_{\ve{\phi}}(\theta) &=  \exdi{z;\theta}{ \ve{\phi}(z) \ve{\phi}^{\T}(z) } - \ve{\mu}_{\ve{\phi}}(\theta) \ve{\mu}^{\T}_{\ve{\phi}}(\theta).
\end{align}
This allows to reformulate the individual parts of \eqref{eq:strong:fbound:unopt} by
\begin{align} \label{reform:nominator}
\sum\limits_{l=1}^{L} \beta_l(\theta)  \frac{\partial \exdi{z;\theta}{\phi_l(z)} }{\partial\theta} = \ve{\beta}^{\T} (\theta) \frac{\partial \ve{\mu}_{\ve{\phi}}(\theta)}{\partial \theta}
\end{align}
and
\begin{align} \label{reform:denominator}
&\exdi{z;\theta}{ \Big( \sum\limits_{l=1}^{L} \beta_l(\theta) \phi_l(z) \Big)^2 } - \Big( \sum\limits_{l=1}^{L} \beta_l(\theta)\exdi{z;\theta}{\phi_l(z)}\Big)^2=\notag\\
&=\ve{\beta}^{\T}(\theta) \ve{R}_{\ve{\phi}}(\theta) \ve{\beta}(\theta).
\end{align}
With \eqref{reform:nominator} and \eqref{reform:denominator} it is possible to state an alternative form of the Fisher information bound \eqref{eq:strong:fbound:unopt} 
\begin{align}\label{bound:unopt:lmoments}
F(\theta)\geq \frac{\ve{\beta}^{\T}(\theta) \frac{\partial \ve{\mu}_{\ve{\phi}}(\theta)}{\partial \theta} \Big(\frac{\partial \ve{\mu}_{\ve{\phi}}(\theta)}{\partial \theta}\Big)^{\T} \ve{\beta}(\theta) }{\ve{\beta}^{\T}(\theta) \ve{R}_{\ve{\phi}}(\theta) \ve{\beta}(\theta)}.
\end{align}
With the substitution
\begin{align}
\ve{\beta}(\theta)= \ve{R}_{\ve{\phi}}^{-\frac{1}{2}}(\theta)  \ve{\beta}'(\theta),
\end{align}
the tightest bound (\ref{bound:unopt:lmoments}) is found by solving the quadratic maximization problem
\begin{align}\label{problem:optim:bound:lmoments}
\max_{\ve{\beta}'(\theta)\in\fieldR^{L}} & \ve{\beta}'^{\T}(\theta) \bigg( \ve{R}_{\ve{\phi}}^{-\frac{1}{2}}(\theta)\frac{\partial \ve{\mu}_{\ve{\phi}}(\theta)}{\partial \theta} \bigg(\frac{\partial \ve{\mu}_{\ve{\phi}}(\theta)}{\partial \theta}\bigg)^{\T}  \ve{R}_{\ve{\phi}}^{-\frac{1}{2}}(\theta) \bigg) \ve{\beta}'(\theta)\notag\\
&\text{s.t. }{\ve{\beta}'^{\T}(\theta) \ve{\beta}'(\theta)=1}
\end{align}
with respect to $\ve{\beta}'(\theta)$.
An optimization problem with the structure of (\ref{problem:optim:bound:lmoments}) is solved by 
\begin{align}
\ve{\beta}'^{\star}(\theta)=\ve{d}_1(\theta),
\end{align}
where $\ve{d}_1(\theta)$ is the eigenvector of the matrix 
\begin{align}\label{matrix:d}
\ve{D}(\theta)=\ve{R}_{\ve{\phi}}^{-\frac{1}{2}}(\theta)\frac{\partial \ve{\mu}_{\ve{\phi}}(\theta)}{\partial \theta} \bigg(\frac{\partial \ve{\mu}_{\ve{\phi}}(\theta)}{\partial \theta}\bigg)^{\T}  \ve{R}_{\ve{\phi}}^{-\frac{1}{2}}(\theta),
\end{align}
corresponding to the principal eigenvalue $\zeta_1(\theta)$ of $\ve{D}(\theta)$. As here the matrix $\ve{D}(\theta)$ has rank one, it is possible to show that
\begin{align}
\label{eigenvector:principal}
\ve{d}_1(\theta)&=\frac{\ve{R}_{\ve{\phi}}^{-\frac{1}{2}}(\theta)\frac{\partial \ve{\mu}_{\ve{\phi}}(\theta)}{\partial \theta}}{\sqrt{\big(\frac{\partial \ve{\mu}_{\ve{\phi}}(\theta)}{\partial \theta}\big)^{\T}  \ve{R}_{\ve{\phi}}^{-1}(\theta)\frac{\partial \ve{\mu}_{\ve{\phi}}(\theta)}{\partial \theta}}}
\end{align}
and
\begin{align}
\label{eigenvalue:principal}
\zeta_1(\theta)&=\bigg(\frac{\partial \ve{\mu}_{\ve{\phi}}(\theta)}{\partial \theta}\bigg)^{\T}  \ve{R}_{\ve{\phi}}^{-1}(\theta)\frac{\partial \ve{\mu}_{\ve{\phi}}(\theta)}{\partial \theta}.
\end{align}
\begin{theorem}[Matched Fisher Information Bound]
For a probability density or mass function $p(z;\theta)$ and any set of $L$ deterministic output transformations $\ve{\phi}(z)$, with the definitions \eqref{mean:aux:stat} and \eqref{covariance:aux:stat}, the Fisher information measure $F(\theta)$ is lower bounded by
\begin{align}\label{bound:eigenvalue}
F(\theta)\geq\bigg(\frac{\partial \ve{\mu}_{\ve{\phi}}(\theta)}{\partial \theta}\bigg)^{\T}  \ve{R}_{\ve{\phi}}^{-1}(\theta)\frac{\partial \ve{\mu}_{\ve{\phi}}(\theta)}{\partial \theta}.
\end{align}
\end{theorem}
\begin{IEEEproof}
Follows from the fact that with the solution of \eqref{problem:optim:bound:lmoments} the right hand side of \eqref{bound:unopt:lmoments} obtains the value given in \eqref{eigenvalue:principal}. 
\end{IEEEproof}
\section{Quality of Nonlinear Systems}
In order to demonstrate the application of the information bound \eqref{bound:eigenvalue} for systems with unknown or difficult analytic model description $p(z;\theta)$, we use three nonlinear examples. For each of them we use $L=7$ and
\begin{align}\label{example:output:trans}
\ve{\phi}(z)=
\begin{bmatrix}
z &z^2 &z^3 &z^4 &|z| &\ln |z| &\ln^2 |z|
\end{bmatrix}^{\T}
\end{align}
with the aim to obtain a measurement-driven approximation of the Fisher information based on the considered output transformations. 
\begin{table}[!htbp]
\renewcommand{\arraystretch}{1.3}
\caption{Exponential Family Distributions}
\label{tab:exp:dist}
\centering
\begin{tabular}{l||c||c}
\hline
\bfseries Distribution & \bfseries $t_1(z)$ & \bfseries $t_2(z)$\\
\hline\hline
Bernoulli & $z$ & -\\
Binomial & $z$ & -\\
Chi-squared & $\ln z$ & -\\
Dirichlet & $\ln z$ & -\\
Exponential & $z$ & -\\
Gaussian & $z$ & $z^2$\\
Gamma & $\ln z$ & $z$\\
Laplace (zero-mean) & $|z|$ & -\\
Log-Normal & $\ln z$ & $\ln^2 z$\\
Poisson & $z$ & -\\
\hline
\end{tabular}
\end{table}
Our choice \eqref{example:output:trans} is motivated by the common distributions within the exponential family (see Tab. \ref{tab:exp:dist}) and their sufficient statistics. We expect that intractable practical distributions may be some mixture of these common probability laws. In such a case one can use a linear combination of the sufficient statistics in order to reach a proper bound. With $N=10^9$ realizations for each calibrated setup of $\theta$, the required mean and covariance of the transformations are approximated by their sample mean
\begin{align}\label{mean:aux:stat:approx}
\ve{\mu}_{\ve{\phi}}(\theta)\approx \frac{1}{N} \sum_{n=1}^{N}\ve{\phi}(z_n)
\end{align}
and
\begin{align}\label{covariance:aux:stat:approx}
\ve{R}_{\ve{\phi}}(\theta) &\approx  \Big( \frac{1}{N} \sum_{n=1}^{N}\ve{\phi}(z_n) \ve{\phi}^{\T}(z_n) \Big) \notag\\
&- \Big(\frac{1}{N} \sum_{n=1}^{N}\ve{\phi}(z_n)\Big) \Big(\frac{1}{N} \sum_{n=1}^{N}\ve{\phi}(z_n)\Big)^{\T}.
\end{align}
After numerically approximating the derivative of $\ve{\mu}_{\ve{\phi}}(\theta)$ it is possible to calculate a pessimistic approximation of the Fisher information measure by the lower bound \eqref{bound:eigenvalue} and to state a conservative version of the CRLB \eqref{bound:crlb}.
\subsection{Saleh model}
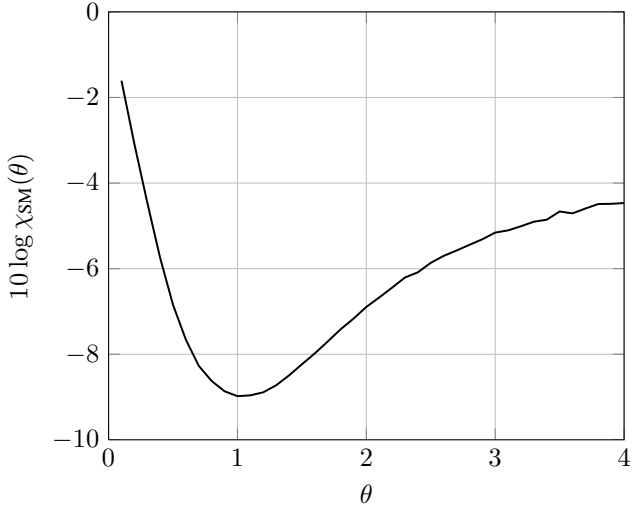
\begin{figure}[!htbp]
\begin{tikzpicture}[scale=1.0]

  	\begin{axis}[ylabel=$10 \log{\chi_{\text{SM}}(\theta)}$,
  			xlabel=$\theta$,
			grid,
			ymin=-10.0,
			ymax=0,
			xmin=0,
			xmax=4.0,
			legend pos=south west]
			
    			\addplot[black, style=solid, line width=0.75pt] table[x index=0, y index=1]{SalehPower.txt};
						
\end{axis}	
\end{tikzpicture}
\caption{Information Loss - Saleh model ($a=2.1587, b=1.1517$)}
\label{Saleh_Loss}
\end{figure}

The first example is the Saleh model
\begin{align}\label{nlmodel:saleh}
z=\frac{a x}{1+b x^2},
\end{align}
where $a,b\in\fieldR$ are model parameters. This nonlinear model was originally developed in order to characterize amplitude saturation effects in traveling-wave-tube amplifiers \cite{Saleh81}. We use the developed approximation \eqref{bound:eigenvalue} in order to investigate the quality of \eqref{nlmodel:saleh} with respect to the problem of measuring the system input power. To this end, we assume that the input $x$ of the nonlinear model follows a Gaussian distribution, i.e., $x\sim\mathcal{N}(0,\theta)$ with $\theta>0$. Fig. \ref{Saleh_Loss} shows the information loss 
\begin{align}\label{loss:saleh}
\chi_{\text{SM}}(\theta)&=\frac{1}{F_x(\theta)}\bigg(\frac{\partial \ve{\mu}_{\ve{\phi}}(\theta)}{\partial \theta}\bigg)^{\T}  \ve{R}_{\ve{\phi}}^{-1}(\theta)\frac{\partial \ve{\mu}_{\ve{\phi}}(\theta)}{\partial \theta}
\end{align}
in $\text{dB}$ with the setup $a=2.1587, b=1.1517$. Note that the Fisher information measure  with respect to $\theta$ at the system input $x$ is $F_x(\theta)=\frac{1}{2\theta^2}$.
It can be observed that the Saleh model \eqref{nlmodel:saleh}, under the considered choice of $a$ and $b$, performs good for small and large input power values $\theta$. For unit input variance a substantial information loss, in comparison to the input, is present at the output of the amplifier.
\subsection{Rician model}
\begin{figure}[!htbp]
\begin{tikzpicture}[scale=1.0]

  	\begin{axis}[ylabel=$10 \log{\chi_{\text{RM}}(\theta)}$,
  			xlabel=$\theta$,
			grid,
			ymin=-25.0,
			ymax=0,
			xmin=0,
			xmax=2.0,
			legend pos=south west]
			
    			\addplot[black, style=solid, line width=0.75pt, smooth] table[x index=0, y index=1]{Rice.txt};
						
\end{axis}	
\end{tikzpicture}
\caption{Information Loss - Rician model}
\label{Rician_Loss}
\end{figure}
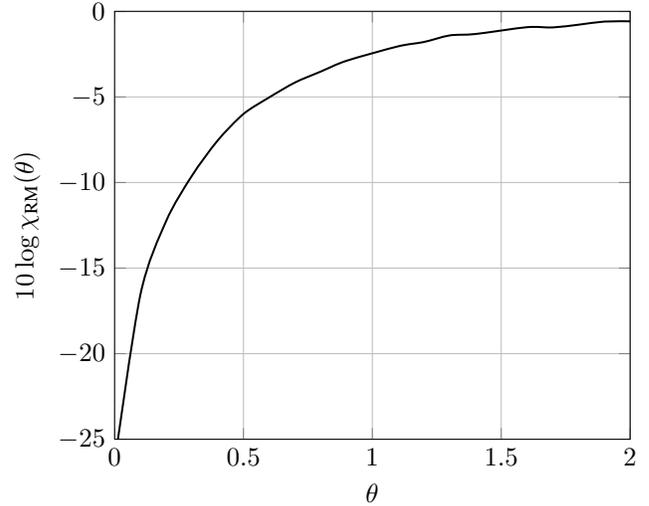

As a second example we investigate a Rician model
\begin{align}\label{model:rice}
z=\sqrt{x_1^2+x_2^2},
\end{align}
where $x_1\sim\mathcal{N}(\theta \cos(a),1)$ and $x_2\sim\mathcal{N}(\theta \sin(a),1)$. Such a model is popular in wireless communications in order to describe the strength of the line-of-sight (LOS) propagation in relation to multi-path channels\cite{Medawar13,Tepe03}. Further, one finds such models in biomedical signal processing for the characterization of brain scan images with fMRI \cite{Barbe11}. Fig. \ref{Rician_Loss} shows the information loss of the Rician system
\begin{align}\label{loss:rice}
\chi_{\text{RM}}(\theta)&=\frac{1}{F_x(\theta)}\bigg(\frac{\partial \ve{\mu}_{\ve{\phi}}(\theta)}{\partial \theta}\bigg)^{\T}  \ve{R}_{\ve{\phi}}^{-1}(\theta)\frac{\partial \ve{\mu}_{\ve{\phi}}(\theta)}{\partial \theta}
\end{align}
in $\text{dB}$. Note, that the Fisher information measure with respect to $\theta$ under direct access to both inputs $x_1$ and $x_2$ is $F_x(\theta)=1$.
It becomes visible that phase information is extremely important for small values of the distance parameter $\theta$. This kind of information is discarded by the Rician model \eqref{model:rice}.
\subsection{Cubic polynominal model}
\begin{figure}[!htbp]

\begin{tikzpicture}[scale=1.0]

  	\begin{axis}[ylabel=$\operatorname{NRMSE}_{\text{CM}}(\theta)$,
  			xlabel=$\theta$,
			grid,
			ymin=0.0,
			ymax=15,
			xmin=0.1,
			xmax=1.0,
			legend pos=north east]
			
    			\addplot[black, style=solid, line width=0.75pt, every mark/.append style={solid}, mark=diamond*, mark repeat=2] table[x index=0, y index=1]{CubicReg_1_0.txt};
			\addlegendentry{$a=0, b=1$}
			
			\addplot[black, style=dashed, line width=0.75pt, every mark/.append style={solid}, mark=otimes*, mark repeat=2] table[x index=0, y index=1]{CubicReg_1_1.txt};
			\addlegendentry{$a=1, b=1$}
			
			\addplot[blue, style=solid, line width=0.75pt, every mark/.append style={solid}, mark=square*, mark repeat=2] table[x index=0, y index=1]{CubicReg_2_0.txt};
			\addlegendentry{$a=0, b=2$}
			
			\addplot[blue, style=dashed, line width=0.75pt, every mark/.append style={solid}, mark=triangle*, mark repeat=2] table[x index=0, y index=1]{CubicReg_2_2.txt};
			\addlegendentry{$a=2, b=2$}
						
\end{axis}	
\end{tikzpicture}
\caption{Pessimistic CRLB - Cubic polynomial model}
\label{CubicCRLB}
\end{figure}
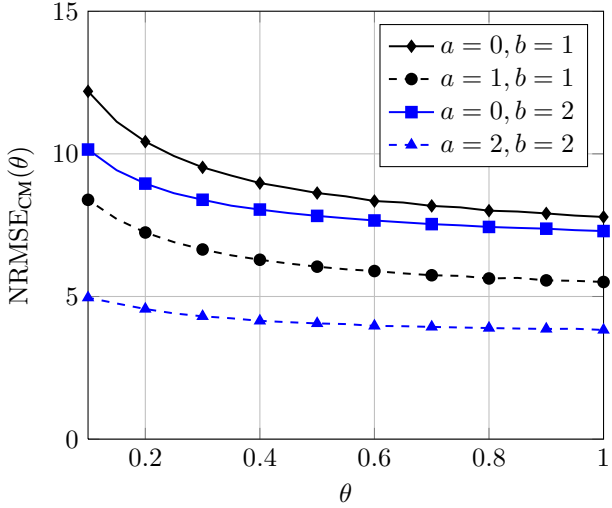

Finally we consider a regression problem and use our approximation \eqref{bound:eigenvalue} in order to determine the smallest possible mean-square error (MSE) for the regression parameter $\theta$. The considered system model is a cubic polynomial
\begin{align}
z=\theta x^3 + x.
\end{align}
We assume a Gaussian input $x\sim\mathcal{N}(a,b)$ and calculate the pessimistic CRLB with respect to the parameter $\theta$
\begin{align}
\text{CRLB}_{\text{CM}}(\theta)= \Bigg(\bigg(\frac{\partial \ve{\mu}_{\ve{\phi}}(\theta)}{\partial \theta}\bigg)^{\T}  \ve{R}_{\ve{\phi}}^{-1}(\theta)\frac{\partial \ve{\mu}_{\ve{\phi}}(\theta)}{\partial \theta}\Bigg)^{-1}.
\end{align}
Fig. \ref{CubicCRLB} shows the conservative approximation of the normalized MSE
\begin{align}
\operatorname{NRMSE}_{\text{CM}}(\theta) = \sqrt{\frac{\text{CRLB}_{\text{CM}}(\theta)}{\theta^2}} 
\end{align}
for different values of the regression parameter $\theta$ and input signal setups $a,b$.
It becomes visible that the configuration of the input signal plays a significant role for the estimation of $\theta$. A higher variance and an input mean different from zero will lead to better performance with an unbiased estimator of the regression parameter $\theta$.
\section{Conclusion}
We have presented a conservative framework which allows to determine the quality of nonlinear stochastic systems in a measurement-driven way. By replacing the original system with an equivalent counterpart which belongs to the exponential family, we showed that it is possible to derive a lower bound for the Fisher information measure. All entities which are required to determine the adequate replacement model can be obtained by calibrated measurements of the system output. This allows to analyze the estimation theoretic quality of measurement systems with complicated or unknown analytical input-to-output relationship. With different examples we have demonstrated that the result is of interest for various applications in signal processing and system identification.
\appendices
\section{Fisher Information Bound}\label{append:fish:bound}
\begin{IEEEproof}
Define
\begin{align}
f(z;\theta)&=\frac{\partial \ln{p(z;\theta)}}{\partial\theta},\label{gen:bound:func1}
\end{align}
and
\begin{align}
g(z;\theta) &= \sum_{l=1}^{L} \beta_l(\theta) \phi_l(z) - \alpha(\theta),
\label{gen:bound:func2}
\end{align}
where $\beta_l(\theta)$ and $\alpha(\theta)$ are weighting factors, free to design. The two functions \eqref{gen:bound:func1} and \eqref{gen:bound:func2} have the properties
\begin{align}
&\int_{\mathcal{Z}} f(z;\theta)g(z;\theta)p(z;\theta) {\rm d}z = \notag\\ 
&=\sum_{l=1}^{L} \beta_l(\theta)\int_{\mathcal{Z}} \frac{\partial {p(z;\theta)} }{\partial\theta}   \phi_l(z)  {\rm d}z - \alpha(\theta) \int_{\mathcal{Z}} \frac{\partial {p(z;\theta)}}{\partial\theta} {\rm d}z\notag\\
&=\sum_{l=1}^{L} \beta_l(\theta)  \frac{\partial \exdi{z;\theta}{\phi_l(z)} }{\partial\theta}
\end{align}
and
\begin{align}
&\int_{\mathcal{Z}} g^2(z;\theta)p(z;\theta) {\rm d}z = \int_{\mathcal{Z}} \bigg( \sum_{l=1}^{L} \beta_l(\theta) \phi_l(z) \bigg)^2 p(z;\theta) {\rm d}z\notag\\
&-2 \alpha(\theta) \sum_{l=1}^{L} \beta_l(\theta)\int_{\mathcal{Z}} \phi_l(z) p(z;\theta) {\rm d}z+\alpha^2(\theta)\notag\\ 
&=\exdi{z;\theta}{ \bigg( \sum_{l=1}^{L} \beta_l(\theta) \phi_l(z) \bigg)^2 } -2 \alpha(\theta) \sum_{l=1}^{L} \beta_l(\theta)\exdi{z;\theta}{\phi_l(z)}\notag\\ 
&+\alpha^2(\theta).
\end{align}
In conjunction with the Cauchy-Schwarz inequality
\begin{align}\label{ineq:cs}
\int_{\mathcal{Z}} {f^2(z;\theta)} p(z;\theta){\rm d}z&\geq \frac{\Big(\int_{\mathcal{Z}} f(z;\theta)g(z;\theta)p(z;\theta) {\rm d}z\Big)^2}{\int_{\mathcal{Z}}g^2(z;\theta)p(z;\theta) {\rm d}z},
\end{align}
a generic lower bound for the Fisher information measure
\begin{align}\label{eq:fbound:generic}
&F(\theta)\geq \frac{\Big(\int_{\mathcal{Z}} f(z;\theta)g(z;\theta)p(z;\theta) {\rm d}z\Big)^2}{\int_{\mathcal{Z}}g^2(z;\theta)p(z;\theta) {\rm d}z}\notag\\ 
&=\frac{\Big( \sum \beta_l(\theta)  \frac{\partial \exdi{}{\phi_l(z)} }{\partial\theta} \Big)^2}{\exdi{}{ \big( \sum \beta_l(\theta) \phi_l(z) \big)^2 } -2 \alpha(\theta) \sum \beta_l(\theta)\exdi{}{\phi_l(z)}+\alpha^2(\theta)}
\end{align}
is obtained. As the optimization problem
\begin{align}
x^\star=\arg \max_{x\in\fieldR} h(x)
\end{align}
with the objective function
\begin{align}
h(x)=\frac{a}{b - 2x c + x^2 }
\end{align}
has a unique solution $x^\star=c$, the strongest form of the bound \eqref{eq:fbound:generic} is obtained by setting
\begin{align}\label{replacement:alpha}
\alpha(\theta)=\sum_{l=1}^{L} \beta_l(\theta)\exdi{z;\theta}{\phi_l(z)}.
\end{align}
Using \eqref{replacement:alpha} in \eqref{eq:fbound:generic} results in \eqref{eq:strong:fbound:unopt}.
\end{IEEEproof}
\bibliographystyle{IEEEbib}

\begin{thebibliography}{20}

\bibitem{Medawar13}
S. Medawar, P. Handel, P. Zetterberg, ``Approximate maximum likelihood estimation of Rician K-factor and investigation of urban wireless measurements," \textit{IEEE Trans. Wireless Commun.}, vol. 12, no. 6, pp. 2545--2555, June 2013.

\bibitem{Oehlert92}
G. W. Oehlert, ``A note on the Delta method",  \textit{Amer. Statisticiany}, vol. 46, no. 1, pp. 27--29, Feb. 1992

\bibitem{Rowe82}
H. E. Rowe, ``Memoryless nonlinearities with Gaussian inputs: Elementary results," \textit{Bell Syst. Tech. J.}, vol. 61, no. 7, pp. 1519--1525, Sept. 1982.

\bibitem{Barbe14}
K. Barb\'e, L. Gonzales Fuentes, L. Barford, L. Lauwers, ``A guaranteed blind and automatic probability density estimation of raw measurements," \textit{IEEE Trans. Instrum. Meas.}, vol. 63, no. 9, pp. 2120--2128, Sept. 2014.

\bibitem{Stoica11}
P. Stoica, P. Babu, ``The Gaussian data assumption leads to the largest Cram{\`e}r-Rao bound," \textit{IEEE Signal Process. Mag.}, vol. 28, pp. 132--133, May 2011.

\bibitem{Stein14}
M. Stein, A. Mezghani, J. A. Nossek, ``A lower bound for the Fisher information measure," \textit{IEEE Signal Process. Letters}, vol. 21, no. 7, pp. 796--799, July 2014.

\bibitem{Saleh81}
 A. A. M. Saleh, ``Frequency-independent and frequency-dependent nonlinear models of TWT amplifiers," \textit{IEEE Trans. Commun.}, vol. 29, no. 11, pp.1715--1720, Nov. 1981.
 
\bibitem{Tepe03}
C. Tepedelenlioglu, A. Abdi, G. B. Giannakis, ``The Ricean K factor: estimation and performance analysis," \textit{IEEE Trans. Wireless Commun.}, vol. 2, no. 4, pp. 799--810, July 2003.

\bibitem{Barbe11}
K. Barb\'{e}, W. Van Moer, L. Lauwers, ``Functional Magnetic Resonance Imaging: An improved short record signal model," \textit{IEEE Trans. Instrum. Meas.}, vol. 60, no. 5, pp. 1724--1731, 2011.

\bibitem{Bai03}
E.-W. Bai, ``Frequency domain identification of Wiener models", \textit{Automatica}, vol. 39, no. 9, , pp. 1521--1530, Sept. 2003

\bibitem{Kay93}
S. M. Kay, \textit{Fundamentals of Statistical Signal Processing: Estimation Theory}. Upper Saddler River, NJ: Prentice Hall, 1993.

\bibitem{Rao45}
C. R. Rao, ``Information and accuracy attainable in the estimation of statistical parameters," \textit{Bull. Calcutta Math. Soc.}, vol. 37, no. 3, pp. 81--91, 1945.

\bibitem{Cram46}
H. Cram\'er, \textit{Mathematical Methods of Statistics}. Princeton, NJ: Princeton Univ. Press, 1946.

\bibitem{SteinWCL14}
M. Stein, S. Theiler, J. A. Nossek, ``Overdemodulation for high-performance receivers with low-resolution ADC,'' \textit{IEEE Wireless Commun. Letters}, vol. 4, no. 2, pp.169--172, 2015.

\end{thebibliography}

\end{document}